\documentclass{elsarticle}
\usepackage{subfigure}
\usepackage{rotating}
\usepackage{lscape}
\usepackage[table]{xcolor}
\usepackage{longtable}
\usepackage{makecell}
\usepackage{siunitx}
\usepackage{gensymb}
\usepackage{natbib}
\biboptions{comma,round}
\setcitestyle{authoryear}
\usepackage{hyperref}

\begin{document}

\title{State capacity and vulnerability to natural disasters}

\author[R.S.J. Tol]{Richard S.J. Tol}

\address{Department of Economics, University of Sussex, Falmer, Jubilee Building, BN1 9SL, UK\\
Institute for Environmental Studies, Vrije Universiteit, Amsterdam, The Netherlands\\
Department of Spatial Economics, Vrije Universiteit, Amsterdam, The Netherlands\\
Tinbergen Institute, Amsterdam, The Netherlands
CESifo, Munich, Germany\\
Payne Institute for Earth Resources, Colorado School of Mines, Golden, CO, USA\\
r.tol@sussex.ac.uk}

\begin{abstract}
Many empirical studies have shown that government quality is a key determinant of vulnerability to natural disasters. Protection against natural disasters can be a public good\textemdash flood protection, for example\textemdash or a natural monopoly\textemdash early warning systems, for instance. Recovery from natural disasters is easier when the financial system is well-developed, particularly insurance services. This requires a strong legal and regulatory environment. This paper reviews the empirical literature to find that government quality and democracy reduce vulnerability to natural disasters while corruption of public officials increases vulnerability. The paper complements the literature by including tax revenue as an explanatory variable for vulnerability to natural disasters, and by modelling both the probability of natural disaster and the damage done. Countries with a larger public sector are better at preventing extreme events from doing harm. Countries that take more of their revenue in income taxes are better that reducing harm from natural disasters.
\end{abstract}

\maketitle

\section{Introduction}
Poorer countries are generally found to be more vulnerable to weather extremes \citep{Cavallo2011} and to climate change \citep{Tol2018}. This is typically explained by the structure of underdeveloped economies\textemdash agriculture is most exposed to the weather\textemdash their geographic location\textemdash high temperatures, heavy downpours, violent storms\textemdash and the lack of adaptive capacity \citep{ADGER2006, Yohe2002}. Adaptative capacity is an aggregate term, comprising (\textit{i}) the availability of protective technologies such as dikes and storm shelters, (\textit{ii}) the economic resources to pay for those technologies, (\textit{iii}) the political will to mobilize those resources, and (\textit{iv}) the ability of government to deliver often complex projects. As an historical illustration of the last two points, the Netherlands began its now-famed dike building programme shortly after the 1849 constitution established a strong central government that answered to a broad share of the population \citep{TolLangen2000}. Because adaptive capacity is a catchall term, it provides little guidance for policy advice on strategies to reduce vulnerability. Most studies proxy adaptive capacity by per capita income and parameterize it with an income elasticity \citep{burke2016adaptation}. The policy recommendation that follows is to grow rich before the next natural disaster hits.

This paper focuses on \emph{state capacity}, a key component of adaptive capacity. State capacity is the ability of the government to raise taxes and provide public goods \citep{Besley2009}. Although a key part of adaptive capacity, it is often overlooked and imperfectly understood in the disaster literature. Governments play a key role in reducing vulnerability, including building dikes and shelters to protect property and lives against floods, providing extension services to make farms more robust to weather extremes, building dams and canals for irrigation to protect against droughts, setting building standards to mitigate storm and earthquake damage, and improving sanitation to ward against the diseases, such as diarrhea and cholera, that often follow in the wake of a natural disaster. These examples are all either (impure) public goods and services, or provided by the government in many if not most countries. Therefore, a lack of ability to raise taxes and a limited capability to deliver public projects hamper protection against extreme weather and so increase vulnerability.

The disaster literature has paid attention to corruption of public officials. This is an important reason for government failure to deliver, and has been found to correlate with vulnerability to natural disasters \citep{Escaleras2007, Escaleras2016}. So has institutional quality \citep{Kahn2005, Persson2017}. \citet{Toya2007, Lin2015a, Lim2019, Fankhauser2014} and \citet{Benali2019} study the relationship between public spending and vulnerability to natural disasters, but the ability to raise taxes has yet to be explored. This paper does that.

This paper also advances the literature by considering both the intensive and the extensive margin of vulnerability. I study both the harm done by natural disasters\textemdash as is common\textemdash and\textemdash less common\textemdash the probability of a natural disaster happening (for one data-set) or the probability of an extreme event doing damage (for another data-set). I find that reducing the probability of being harmed matters more than reducing the harm. 

The paper proceeds as follows. Section \ref{sc:disasters} reviews the rapidly growing literature on the correlates of natural disasters. Section \ref{sc:method} presents the data and methods used to analyze them. Section \ref{sc:results} discusses the results. Section \ref{sc:conclude} concludes.

\section{Previous literature}
\label{sc:disasters}

\subsection{The impact of natural disasters on development}
Natural disasters have a number of effects on the economy \citep{Albala1993}. When disaster strikes, economic activity is disrupted and input factors are destroyed. Some disasters, such as floods and storms, particularly affect physical capital, such as buildings, roads and bridges. Other disasters, such as epidemics, primarily affect people and thus the current or future labour force. After the disaster, there is recovery. The dead are buried, debris cleared away. Savings are mobilized, and insurance payouts and charity received to rebuild houses and roads. These are economic activities, and so contribute to Gross Domestic Product (GDP). Natural disasters thus neatly illustrate the \textit{broken window fallacy} of \citet{Bastiat1850}: Destruction of the capital stock is not measured by GDP (but rather by Net Domestic Product). Repair of the capital stock is included in GDP. A naive look at GDP growth rates may therefore lead one to conclude that natural disasters are good for short-term economic growth. This is not the case. Natural disasters stimulate economic activity. Natural disasters do not improve welfare. GDP is a measure of economic activity. It is not a welfare indicator. In cases like these, we should focus on NDP growth rates, but data availability is limited.

Natural disasters have different impacts at different phases of the business cycle \citep{TolLeek1998}. During a recession, the loss of input factors is less problematic as there is overcapacity anyway. The recovery phase, if it materializes, is a Keynesian economic stimulus. During a boom, capacity is tight and lost inputs cannot readily be replaced. The demand stimulus from recovery may drive up inflation rather than output. Recovery replaces destroyed capital goods with new ones. Although the initial response is often to restore things \emph{exactly as they were}, in fact replacements are often superior: New machinery would be state-of-the-art, new buildings better designed, new roads without previous bottlenecks, and so on. This does not accelerate economic growth in the long run: The capital stock would be replaced anyway. Natural disasters force the hand of economic agents with regard to the timing of replacement investment. Discretionary timing would be preferred. There is a chance that a natural disaster destroys an entrenched monopoly or topples a corrupt government\textemdash this would accelerate economic growth\textemdash but a natural disaster is just as likely to entrench corruption or destroy market challengers.

The impact of natural disasters on the economy in the short term is therefore mixed, but probably negative on net. The same is true in the long term. If there is a risk of natural disasters, investment is diverted from productive to defensive capital, such as dykes. The return on defensive investment is essentially zero. If the disaster risk were lower, resources could be used for consumption or for investment in productive assets. The risk of natural disasters also increases the demand for reserves and insurance, invested in liquid assets with a low return. As with defensive investment, the long-run economic impact of natural disasters is not their cost, but their opportunity cost. 

Natural disasters have different impacts on different economies \citep{Cavallo2011, Kellenberg2011}. Recovery requires resources. In developed economies, recovery is paid for by insurance, from household and company reserves, by the government, or by new loans from commercial lenders. In developing economies, contributions from these sources are limited, and recovery depends on support from family, informal networks, and charity. Recovery from natural disasters is therefore slower, and sometimes much slower in developing countries than in developed economies. There may be hysteresis if customers switch to new suppliers. For instance, it takes several years to establish a banana plantation. Bananas are often sold on multi-year contracts. A new contract need not be on offer for a plantation restored after a hurricane.

There is now a substantial body of literature that shows that natural disasters indeed slow down economic growth \citep{Klomp2014meta, Lazzaroni2014, Banica2020, Parida2021}. This negative effect is stronger in poorer countries, partly because financial underdevelopment hampers recovery. Financial underdevelopment may itself be caused by natural disasters \citep{McDermott2013, Klomp2014, Klomp2015, Keerthiratne2017, Klomp2017, Klomp2018, An2019, Klomp2020, Horvath2021}. That is, natural disasters slow economic growth and more so in poorer countries while poorer countries are more vulnerable to natural disasters. In other words, areas which are more prone to extreme weather are more likely to be trapped in poverty.

\subsection{The impact of development on natural disasters}
The empirical evidence on the determinants of the impact of natural disasters has grown rapidly, revealing heterogeneity in many dimensions. See Table \ref{tab:vulnerable}. \citet{Kahn2005} is often seen as the starting point of research into the correlates of vulnerability to natural disasters, but \citet{Yohe2002} were first. These and many other studies use the EM-DAT data,\footnote{See \url{https://public.emdat.be/}.} and it was the decision to make this data available for research that was impetus for this branch of literature. Before that, the literature was thin and either qualitative or formal.\footnote{There is also a large literature in social geography, mostly on case studies of natural disasters.}

Unfortunately, EM-DAT only reports natural disasters that cause substantial damage to humans. This compounds the endogeneity: Natural disasters hamper economic growth and richer societies, better able to shield themselves from natural disasters, are less likely to record natural disasters. The data by \citet{FELBERMAYR2013} does not suffer this problem, reporting also storms, floods, droughts, earthquakes and volcanoes that did little damage.\footnote{See \url{https://www.ifo.de/en/ebdc-data/ifo-game}.} Although that data has been available for a while, it is used less often.

Table \ref{tab:vulnerable} lists studies of the correlates of vulnerability to natural disasters. Poverty is the common theme. Poor societies are hit harder. Gender, age, and education feature too.

Institutions are the next biggest topic in this literature. Badly governed countries are more vulnerable to natural disasters. It is generally assumed that that is the direction of causality but \citet{Benali2019} argue that causality runs from natural disasters to public expenditure.

Table \ref{tab:vulnerable} highlights the studies that have some indicator of the quality of public administration. \citet{Kahn2005} uses indicators of democracy, regulatory quality, voice and accountability, rule of law, and control of corruption, \citet{Persson2017} a subset of these. \citet{Zuo2017} add government effectiveness to the mix, and \citet{Enia2018} contract intensity. \citet{Escaleras2007} use public sector corruption. \citet{FELBERMAYR2014} use democracy, \citet{Breckner2016} civil liberties and political rights. \citet{Escaleras2016} use democracy and fiscal decentralization. \citet{Toya2007}, \citet{Fankhauser2014}, \citet{Lin2015a} and \citet{Lim2019} come closest to the current paper, using the size of government, measured as the share of government expenditures to total GDP.\footnote{\citet{Yonson2018} use subnational data on governance for the Philippines. \citet{Iwata2014} have data on local public expenditure on disaster preparedness in Japan, \citet{Liu2018} in China, \citet{Karim2020} in Bangladesh. \citet{Villagra2017} collect data on subnational disaster governance for Chile, \citet{Masiero2020} on post-disaster municipal finance in Italy.} \citet{Toya2007}, \citet{Lin2015a} and \citet{Lim2019} find that government spending reduces the vulnerability to natural disasters, but this effect is insignificant in \citet{Fankhauser2014}.

\begin{landscape}
\begin{footnotesize}
\begin{longtable}{ l c c c }
\caption{Studies of vulnerability to natural disasters.}
\label{tab:vulnerable}
\cr
\hline
study & hazard & region & explanatory variables \\ \hline
\citet{Yohe2002} & natural disasters & world & poverty (+), inequality (+), population density (+) \\
\rowcolor{yellow} \citet{Kahn2005} & natural disasters & world & poverty (+), democracy (-), institutions (-) \\
\rowcolor{yellow} \citet{Toya2007} & natural disasters & world & poverty (+), institutions (-) \\
\rowcolor{yellow} \citet{Escaleras2007} & earthquakes & world & corruption (+) \\
\citet{Kellenberg2008} & natural disasters & world & poverty (- then +) \\
\citet{Price2008} & hurricanes & USA & poverty (+), black (+) \\
\rowcolor{yellow} \citet{Raschky2008} & natural disasters & world & poverty (- then +), institutions (-) \\
\citet{Zhang2009} & drought & China & poverty (+) \\
\citet{Yamamura2010} & natural disasters & Japan & poverty (+), social capital (-) \\
\citet{Keefer2011} & earthquakes & world & poverty (+), democracy (-) \\
\citet{Escaleras2012} & natural disasters & world & decentralization (-) \\
\citet{Hu2012} & earthquakes & China & poverty (+) \\
\citet{Rubin2012} & natural disasters & Latin America & poverty (+), inequality (+) \\
\citet{Wamsler2012} & natural disasters & Brazil, El Salvador & education (-) \\
\citet{Ward2012} & natural disasters & world & poverty (+) \\
\citet{Yamamura2012} & natural disasters & world & ethnic polarization (+) \\
\citet{Chen2013} & natural disasters & China & poverty (+) \\
\rowcolor{yellow} \citet{Ferreira2013} & floods & world & poverty (+), institutions (?) \\
\citet{Ji2013} & natural disasters & world & poverty (+), medical services (-) \\
\citet{Padli2013} & natural disasters & world & poverty (+), education (-) \\
\citet{Skidmore2013} & natural disasters & world & decentralization (-) \\
\rowcolor{yellow} \citet{Fankhauser2014} & hurricanes, floods & world & poverty (+), institutions (?) \\
\rowcolor{yellow} \citet{FELBERMAYR2014} & natural disasters & world & poverty (?), institutions (+), trade (+) \\
\citet{Huang2014} & floods & China, Japan & poverty (- then +) \\
\rowcolor{yellow} \citet{Iwata2014} & natural disasters & Japan & institutions (+) \\
\citet{Muttarak2014} & natural disasters & case studies & education (-) \\
\citet{Rubin2014} & natural disasters & Vietnam & poverty (+), inequality (+) \\
\citet{Zhang2014} & natural disasters & China & poverty (+) \\
\citet{Zhou2014} & natural disasters & China & poverty (+), education (-), young (+), old (+) \\
\citet{Jongman2015} & floods & world & poverty (+) \\
\citet{Lim2015} & heat waves & East Asia & poverty (+) \\
\citet{Lin2015} & earthquakes & Taiwan & poverty (+), female (+), young (+), old (+) \\
\rowcolor{yellow} \citet{Lin2015a} & natural disasters & world & democracy (-), institutions (-) \\
\citet{Park2015} & hurricanes & South Korea & poverty (+), building codes (-) \\
\citet{Toya2015} & natural disasters & world & ICT (-) \\
\citet{Wen2015} & natural disasters & world & right-wing government (-) \\ 
\citet{Austin2016} & natural disasters & world & poverty (+), democracy (-), women's rights (-) \\
\rowcolor{yellow} \citet{Breckner2016} & natural disasters & world & insurance (-), institutions (-) \\
\citet{Dresser2016} & hurricanes & Atlantic & poverty (+) \\
\rowcolor{yellow} \citet{Escaleras2016} & natural disasters & world & corruption (+)\footnote{Escaleras and Register further find that natural disasters increase corruption.} \\
\citet{Klomp2016} & natural disasters & world & poverty (+) \\
\citet{Peregrine2017} & natural disasters & ancient societies & democracy (-) \\
\rowcolor{yellow} \citet{Persson2017} & natural disasters & world & democracy (-), institutions (-) \\
\citet{Tselios2017} & natural disasters & world & decentralization (+) \\
\rowcolor{yellow} \citet{Villagra2017} & tsunami & Chile & institutions (-) \\
\citet{Ward2017} & natural disasters & world & poverty (+) \\
\rowcolor{yellow} \citet{Zuo2017} & natural disasters & world & institutions (-) \\
\citet{Albu2018} & floods & Rio de Janeiro & urban infrastructure (-) \\
\rowcolor{yellow} \citet{Enia2018} & natural disasters & world & rule of law (-) \\
\citet{Hu2018} & floods & world & population density (+), poverty (+) \\
\rowcolor{yellow} \citet{Liu2018} & earthquakes & China & public infrastructure (-), insurance (-) \\
\citet{Peregrine2018} & natural disasters & ancient societies & democracy (-) \\
\citet{Winsemius2018} & floods and droughts & world & poverty (+) \\ 
\citet{Wu2018} & natural disasters & China & poverty (+) \\
\rowcolor{yellow} \citet{Yonson2018} & hurricanes & Philippines & poverty (+), institutions (-), urbanization (+) \\
\rowcolor{yellow} \citet{Fabian2019} & earthquakes & world & poverty (+), institutions (-) \\
\citet{Formetta2019} & natural disasters & world & wealth (-) \\
\citet{Lim2019} & floods & USA & poverty (+), education (-), housing quality (-), local government spending (-) \\
\citet{Miao2019} & floods & world & experience (-) \\
\rowcolor{yellow} \citet{Padli2019} & natural disasters & world & poverty (+), education (-), corruption (+) \\
\rowcolor{yellow} \citet{Parida2019} & floods & India & poverty (+), political alignment (-) \\
\citet{Shen2019} & natural disasters & world & GDP (+), population (+), area (+) \\
\citet{Song2019} & natural disasters & world & GDP (+), population (+), area (+) \\
\citet{Wu2019} & weather disasters & China & poverty (+) \\
\citet{Yang2019} & heat waves & China & poverty (+) \\
\citet{Bahinipati2020} & floods & India & development (?), disaster reduction (-) \\
\citet{DeOliveira2020} & natural disasters & Cear\'{a} & infrastructure (-), population density (-), fiscal decentralization (-) \\
\citet{DeSilva2020} & floods & Rathnapura & poverty (+) \\
\rowcolor{yellow} \citet{Tennant2020} & natural disasters & world & poverty (+), institutions (-) \\
\citet{Klomp2020b} & natural disasters & world & central bank independence (+) \\
\citet{Llorente2020} & earthquakes & Haiti & minority (+), female (+), lower class (+) \\
\citet{Miao2020} & natural disasters & USA & fiscal decentralization (+) \\
\rowcolor{yellow} \citet{Tselios2020} & natural disasters & world & poverty (+), population density (+), education (-), governance (?) \\
\citet{Yabe2020} & hurricanes & Florida & poverty (+) \\
\citet{Yadava2020} & lightning & India & population density (+) \\
\citet{Crowley2021} & hurricanes & Carolinas & minority (+), female (+), elderly (+), young kids (+) \\
\citet{Nguyen2021} & earthquakes & New Zealand & insurance (-) \\
\citet{Nohrstedt2021} & natural disasters & world & poverty (+), disaster reduction (?) \\
\citet{Parida2021b} & natural disasters & Odisha & poverty (+), education (-), infrastructure (-), finance (-) \\
\citet{RoyChowdhury2021} & floods & India & human development (-), female (+) \\
\hline
\end{longtable}
\end{footnotesize}
\end{landscape}

\section{Methods and data}
\label{sc:method}
\subsection{Data}
The key assumption in the model of \citet{Besley2009} is that there is a fixed cost in tax collection, which makes it uneconomic to collect income taxes from the poorest part of the population. The share of people outside the income tax net is larger in poorer countries, and in countries with a larger share of poor people. The World Bank publishes data, per country and year, on total tax revenue relative to the Gross Domestic Product. The same database has income taxes, taxes on international trade, taxes on goods and services, and other taxes\textemdash all as a share of total tax revenue. Composition matters because some taxes are more easily raised (e.g., stamp duties) than others (e.g., income taxes), and because some countries have large non-tax revenues, such as dividends from state-owned oil companies.

The World Bank also publishes data on public expenditure, but spending on keeping people safe from natural disasters is unhelpfully grouped under ``other spending'' with a great many other things. The World Bank also publishes data on population size, corruption, and per capita income, in dollars per person per year, using both market exchange rates and purchasing power parity ones; the latter are also referred to as international dollars. Data on democracy and autocracy are taken from the Polity IV databases.

The dependent variables are the number of people affected and killed by natural disasters, both together and separate for floods, storms, and so on. These data are published, by country and year, by EM-DAT. The CESifo GAME database complements the EM-DAT data with observations on the severity of earthquakes, volcanoes, storms, heatwaves, floods, and droughts, as measured by the Richter scale, the Volcanic Explosivity Index, maximum wind speed, temperature, rainfall, and lack of rainfall, respectively.

\subsection{Methods}
I estimate three kinds of model. The second one is a log-linear model, conditional on an extreme event having occurred (for the CESifo GAME data) or a natural disaster having been reported (for the EM-DAT data). In these models, the dependent variable is the natural logarithm of either the number of people killed or affected, divided by the size of the population in the year and country that the disaster struck. These transformations imply that the data are meaningfully censored from neither below nor above. This is the \emph{intensive} margin, the severity of natural disasters.

The first model is a logit. For the EM-DAT data, the dependent variable equals zero if no disaster is reported, and one if there is. For the GAME data, the dependent variable equals zero if natural scientists recorded an event but EM-DAT did not, and one if both reported an event. This is the \emph{extensive} margin, the occurrence of harmful extreme events.

I estimate the models using Stata. I use random effects panel estimators unless the Hausman test suggest I should use fixed effects instead. I do not use Heckman models, which jointly estimate occurrence and severity, because the panel version has difficulties with convergence. Instead, I check whether the Heckman correction applies.

The third model is a Tobit, only applied to the CESifo GAME data. Observations of extreme weather events without damage are treated as censored observations. Otherwise the model is linear. This model combines the extensive and intensive margins.

\section{Results}
\label{sc:results}

\subsection{EM-DAT}
Table \ref{tab:dis} shows the regression results for the probability of being hit by a natural disaster according to the EM-DAT data. The model used is a panel logit with country random effects. There are four sets of results: a general specification with many explanatory variables (and few observations) and a specific one with insignificant explanatory variables removed; and a specification with per capita income measured in international Geary-Khamis dollars and one using market exchange rates.

Richer countries suffer fewer natural disasters, and democratic countries more. The first result is in line with previous studies, the second is the opposite of what was found before. See Table \ref{tab:vulnerable}. Results on state capacity are mixed. Corruption is not significant, again in contrast with previous studies. If market exchange rates are used, the share of taxes raised by income taxes is borderline significant\textemdash and positive. If purchasing power parity exchange rates are used, the empirical results are in line with theory: A larger state is better able to protect its people against natural disasters.

Table \ref{tab:aff} shows the regression results for the number of people affected, conditional on being hit by a natural disaster, following the set-up of Table \ref{tab:dis}. There is weak evidence that democratic countries are hit harder, and weaker evidence that a larger state means that fewer people are hurt. Corruption, public consumption and investment, and the share of income in total taxes are not significant.

Table \ref{tab:kill} shows the regression results for the number of people killed, conditional on being hit by a natural disaster, following the set-up of Tables \ref{tab:dis} and \ref{tab:aff}. Again, democratic countries are hit harder. Richer countries are hit less hard, if wealth is measured in international dollars. Corruption, public consumption and investment, total tax revenue, and the share of income in total taxes are not significant.

The results in Tables \ref{tab:aff} and \ref{tab:kill} are subject to selection bias. This is small, however. The correlation between the residuals of the intensive and extensive margins is small: -0.036 (market exchange dollars) and -0.090 (international dollars) for the number affected, and 0.016 (market exchange dollars) and 0.071 (international dollars) for the number killed.

\subsection{GAME}
The above regressions use the EM-DAT data, which suffer from selection bias: Only extreme events that do substantial damage are reported. The CESifo GAME data are objective, based on readings of seismographs, anemometers, thermometers and rain gauges. These observations are combined with the EM-DAT data on impacts.

Table \ref{tab:logdis} reports six logit regressions of the probability of deaths being reported given an extreme event. I do this separately for the six main hazards\textemdash earthquakes, volcanic eruptions, high winds, heat waves, floods, and droughts\textemdash because it is not possible to compare severity across hazards.

Stronger earthquakes, bigger storms, greater floods significantly increase the chance of fatalities. The same is true for volcanic eruptions, but the coefficient is weakly significant because there are few observations. There is no relationship between monthly rainfall anomaly and drought, because drought is a sustained lack of rain. The impact of temperature anomaly has the wrong sign, but is only weakly significant.

When an extreme event strikes, richer countries are less likely to report fatalities. Countries with larger tax revenues are safer. However, a higher share of income tax increases vulnerability, except for the case of earthquakes. Democracy does not affect four out of six hazards. Democratic countries are safer from earthquakes\textemdash both this result and the income tax one may be due to Japan\textemdash but less safe from heat\textemdash this may be due to the correlation between temperature and voters' rights.

Table \ref{tab:logdis1} repeats the analysis for the probability that people affected are reported. The patterns are much the same as in Table \ref{tab:logdis}. One exception is heat, where no parameter is significant for people affected. Another difference is that income per capita is highly significant for people affected: Richer countries suffer less from drought (or maybe drought-stricken countries are less likely to be rich).

A different picture emerges from Table \ref{tab:regdis}, which considers the natural logarithm of the number of people killed, normalized by population size, given that at least one was. Severity of the extreme events matters for earthquakes only.

Richer countries are more vulnerable to heatwaves. This may be a reporting problem. Countries with a larger public sector are more vulnerable, but countries that collect more of their tax revenue by income taxation are less so. Democracy is insignificant, except that autocracies keep their people safe from storms.

The fraction of people affected, results for which are in Table \ref{tab:regaff}, show a similar pattern. The severity of the extreme event matters for earthquakes and storms. Richer countries are less vulnerable to floods and droughts; the coefficient for heatwaves is still positive but now insignificant. Countries that collect more tax are more vulnerable, unless that tax is in the form of income taxes. Democracy is insignificant.

The exercise is repeated with country fixed effects. Results are shown in Tables \ref{tab:logdis2}, \ref{tab:logdis12}, \ref{tab:regdis2} and \ref{tab:regaff2}. All results collapse, apart from the effects of the severity of the extreme event and the impact of per capita income. There is too little variation over time in the other explanatory variables, particularly those on state capacity, to have a significant effect.

Tables \ref{tab:tobkill} and \ref{tab:tobaff} present a different take on the same question. Above, I split the extensive margin\textemdash the probability of being harmed by an extreme event\textemdash from the intensive margin\textemdash the damage done by a natural disaster\textemdash using logit and ordinary least squares, respectively. I now combine the two, using a Tobit regression, modelling the extreme events that happened but did not cause harm as observations \emph{censored} at zero. The main disadvantage of this approach is that the dependent variable is now the fraction of people killed or affected, rather than its natural logarithm.

Richer countries are less vulnerable to natural disasters, except for earthquakes. Democratic countries are less vulnerable to earthquakes. Countries with higher tax revenues are less vulnerable. A higher share of tax revenue in income taxes increases vulnerability, except for earthquakes. Comparing this to the results above, the Tobit results are more like the logit results than the OLS results. In other words, the extensive margin is more important than the intensive one.

\section{Discussion and conclusion}
\label{sc:conclude}
I set out to test whether state capacity, the ability of the government to raise taxes and provide public goods, affects vulnerability to natural disasters. Previous literature has shown this to be the case for the ability to deliver: The people in countries with lower quality governance or more corrupt public officials are less safe. I complement this with a study of the size and composition of government revenue. The results are mixed. Using the EM-DAT data, which only report extreme events that do harm, neither government revenue and its composition nor government spending significantly affect either the probability of a natural disaster or its impact. Using the higher-quality GAME data, which also report extreme events that did no harm, people in countries with a larger public sector are less likely to be hurt or killed by a natural disaster. However, extreme events that do do harm, do more harm in these countries. That said, countries that raise more of their revenue from income taxes, are better at reducing harm from natural disasters. Censored regressions suggest that the probability of harm is more important than the harm done for overall vulnerability to natural disasters.

The emerging result is that, in countries with a small public sector, extreme events big and small hurt people. As the public sector grows, it offers more protection against the smaller would-be disasters. The average severity of disasters that do harm therefore goes up. States with a greater capacity to raise taxes from income are better at protecting their people against the larger disasters too.

Future research should investigate whether the impact of the higher share of income taxes on the impact of natural disasters is because this is indeed a good proxy for the capacity of the state to protect its people, because direct taxes induce a greater demand from taxpayers to be protected, or because direct taxes tend to be more progressive than indirect taxes. The income distribution is found to have a significant effect on vulnerability in previous studies. Democracy too is significant in this study and previous ones.

Future research should also move away from studying the correlates of vulnerability\textemdash that literature seems saturated, see Table \ref{tab:vulnerable}\textemdash to causal estimates. Synthetic control and difference-in-difference methods could be used to study the impact of government reforms. Fine-grained spatial data could be used for regression discontinuity analysis along international or internal borders. Local government accounts could be used to research the impact of government spending.

The take-away message from the current paper, which should be a major theme of future research, is that vulnerability to natural disasters does not just evolve over time. Rather, vulnerability is shaped by decisions made by people, companies and\textemdash crucially\textemdash their governments.

\bibliography{statecap}

\begin{table}[htbp]\centering
\def\sym#1{\ifmmode^{#1}\else\(^{#1}\)\fi}
\caption{Regression results: Probability of disaster reported.\label{tab:dis}}
\begin{tabular}{l*{4}{c}}
\hline\hline
                    &\multicolumn{1}{c}{MER general}&\multicolumn{1}{c}{MER specific}&\multicolumn{1}{c}{PPP general}&\multicolumn{1}{c}{PPP specific}\\
\hline
Income per capita   &     -0.0385         &      -0.270         &                   0.583         &      -0.171        \\
                    &     (-0.04)         &     (-1.53)         &                 (0.61)         &     (-0.76)            \\
[1em]
Tax revenue         &      0.0612         &                     &      0.0358         &     -0.0823\sym{*}  \\
                    &      (0.35)         &                     &      (0.20)         &     (-2.53)         \\
[1em]
Income taxes        &      0.0294         &      0.0245         &      0.0236         &                     \\
                    &      (0.74)         &      (1.89)         &      (0.62)         &                     \\
[1em]
Public expenditure  &      0.0134         &                     &    -0.00442         &                     \\
                    &      (0.26)         &                     &     (-0.09)         &                     \\
[1em]
Corruption          &       0.600         &                     &       0.627         &                     \\
                    &      (0.72)         &                     &      (0.76)         &                     \\
[1em]
Democracy           &       0.119         &      0.0425         &       0.109         &      0.0943\sym{**} \\
                    &      (1.01)         &      (1.56)         &      (0.93)         &      (2.73)         \\
\hline
Observations        &          86         &        1104         &          86         &         997         \\
\hline\hline
\multicolumn{5}{l}{\footnotesize All models are estimated with random effects.}\\
\multicolumn{5}{l}{\footnotesize \textit{t} statistics in parentheses}\\
\multicolumn{5}{l}{\footnotesize \sym{*} \(p<0.05\), \sym{**} \(p<0.01\), \sym{***} \(p<0.001\)}\\
\end{tabular}
\end{table}

\begin{table}[htbp]\centering
\def\sym#1{\ifmmode^{#1}\else\(^{#1}\)\fi}
\caption{Regression results: ln(Number of people affected).\label{tab:aff}}
\begin{tabular}{l*{4}{c}}
\hline\hline
                    &\multicolumn{1}{c}{MER general}&\multicolumn{1}{c}{MER specific}&\multicolumn{1}{c}{PPP general}&\multicolumn{1}{c}{PPP specific}\\
\hline
Income per capita   &      -1.408         &      -0.149         &                -1.408         &       0.173         \\
                    &     (-0.99)         &     (-0.44)         &                (-0.99)         &      (0.37)         \\
[1em]
Tax revenue         &      -0.412\sym{**} &      -0.114\sym{*}  &      -0.412\sym{**} &      -0.103         \\
                    &     (-4.38)         &     (-2.13)         &     (-4.38)         &     (-1.76)         \\
[1em]
Income taxes        &      0.0656         &                     &      0.0656         &                     \\
                    &      (1.37)         &                     &      (1.37)         &                     \\
[1em]
Public expenditure  &     0.00910         &                     &     0.00910         &                     \\
                    &      (0.24)         &                     &      (0.24)         &                     \\
[1em]
Corruption          &      -0.568         &                     &      -0.568         &                     \\
                    &     (-0.46)         &                     &     (-0.46)         &                     \\
[1em]
Democracy           &       0.430         &      0.0835\sym{*}  &       0.430         &       0.117\sym{*}  \\
                    &      (0.95)         &      (2.02)         &      (0.95)         &      (2.33)         \\
\hline
Observations        &          58         &         655         &          58         &         580         \\
\hline\hline
\multicolumn{5}{l}{\footnotesize All models are estimated with fixed effects and clustered standard errors for countries.}\\
\multicolumn{5}{l}{\footnotesize \textit{t} statistics in parentheses}\\
\multicolumn{5}{l}{\footnotesize \sym{*} \(p<0.05\), \sym{**} \(p<0.01\), \sym{***} \(p<0.001\)}\\
\end{tabular}
\end{table}

\begin{table}[htbp]\centering
\def\sym#1{\ifmmode^{#1}\else\(^{#1}\)\fi}
\caption{Regression results: ln(Number of people killed).\label{tab:kill}}
\begin{tabular}{l c c c c}
\hline\hline
                    &\multicolumn{1}{c}{MER general}&\multicolumn{1}{c}{MER specific}&\multicolumn{1}{c}{PPP general}&\multicolumn{1}{c}{PPP specific}\\
\hline
Income per capita   &       0.366         &      -0.209         &                1.010         &      -1.250\sym{***}  \\
                    &      (0.71)         &     (-1.69)         &                 (-0.87)         &     (-4.31)         \\
[1em]
Tax revenue         &     -0.0647         &                     &     -0.0412         &     -0.0371         \\
                    &     (-0.93)         &                     &     (-0.39)         &     (-1.42)         \\
[1em]
Income taxes        &     -0.0327\sym{*}  &     -0.0105         &     -0.0413         &                     \\
                    &     (-2.01)         &     (-1.35)         &     (-0.65)         &                     \\
[1em]
Public expenditure  &     -0.0651\sym{*}  &                     &    -0.00586         &                     \\
                    &     (-1.97)         &                     &     (-0.18)         &                     \\
[1em]
Corruption          &      -0.109         &                     &       0.271         &                     \\
                    &     (-0.35)         &                     &      (0.20)         &                     \\
[1em]
Democracy           &      0.0382         &      0.0368\sym{**} &      0.0778         &      0.0767\sym{**} \\
                    &      (0.78)         &      (2.78)         &      (0.57)         &      (3.37)         \\
\hline
Observations        &          56         &         670         &          56         &         583         \\
\hline\hline
\multicolumn{5}{l}{\footnotesize Models with market exchange (international) dollars are estimated with random (country fixed) effects.}\\
\multicolumn{5}{l}{\footnotesize All models are estimated with clustered standard errors for countries.}\\
\multicolumn{5}{l}{\footnotesize \textit{t} statistics in parentheses}\\
\multicolumn{5}{l}{\footnotesize \sym{*} \(p<0.05\), \sym{**} \(p<0.01\), \sym{***} \(p<0.001\)}\\
\end{tabular}
\end{table}

\begin{table}[htbp]\centering
\def\sym#1{\ifmmode^{#1}\else\(^{#1}\)\fi}
\caption{Regression results: Probability of people killed.\label{tab:logdis}}
\begin{tabular}{l*{6}{c}}
\hline\hline
                    &\multicolumn{1}{c}{Earthquake}&\multicolumn{1}{c}{Volcanoe}&\multicolumn{1}{c}{Storm}&\multicolumn{1}{c}{Heat}&\multicolumn{1}{c}{Flood}&\multicolumn{1}{c}{Drought}\\
\hline
Severity&       1.813\sym{***}&     1.065\sym{*}     &   0.0477\sym{***}  &     -0.0947\sym{*}     &   0.779\sym{***}&       0.156     \\
&     (15.93)         &     (2.55)        &   (14.84)       &      (-2.16)        &   (8.82)      &      (0.22)    \\
[1em]
Income per capita   &       0.182\sym{*}  &      -0.578         &      -0.519\sym{***}&      -0.367\sym{***}&      -0.174\sym{***}&      -0.797\sym{*}  \\
                    &      (2.12)         &     (-1.26)         &     (-9.03)         &     (-3.82)         &     (-4.22)         &     (-2.43)         \\
[1em]
Tax revenue         &      -0.129\sym{***}&       0.137         &      -0.114\sym{***}&     -0.0820\sym{***}&      -0.161\sym{***}&     -0.0176         \\
                    &     (-6.12)         &      (1.34)         &     (-9.09)         &     (-3.92)         &    (-15.89)         &     (-0.24)         \\
[1em]
Income taxes        &     -0.0198\sym{**} &    -0.00783         &      0.0350\sym{***}&      0.0166\sym{*}  &      0.0162\sym{***}&      0.0652\sym{*}  \\
                    &     (-3.05)         &     (-0.33)         &      (9.00)         &      (2.31)         &      (4.99)         &      (2.25)         \\
[1em]
Democracy           &     -0.0500\sym{**} &       0.314         &     0.00518         &       0.165\sym{***}&      0.0174         &      -0.131         \\
                    &     (-3.08)         &      (1.26)         &      (0.42)         &      (4.37)         &      (1.80)         &     (-1.80)         \\
\hline
Observations        &        5882         &         316         &        7838         &        7835         &        8064         &        8064         \\
\hline\hline
\multicolumn{7}{l}{\footnotesize \textit{t} statistics in parentheses}\\
\multicolumn{7}{l}{\footnotesize \sym{*} \(p<0.05\), \sym{**} \(p<0.01\), \sym{***} \(p<0.001\)}\\
\end{tabular}
\end{table}

\begin{table}[htbp]\centering
\def\sym#1{\ifmmode^{#1}\else\(^{#1}\)\fi}
\caption{Regression results: Probability of people affected.\label{tab:logdis1}}
\begin{tabular}{l*{6}{c}}
\hline\hline
                    &\multicolumn{1}{c}{Earthquake}&\multicolumn{1}{c}{Volcanoe}&\multicolumn{1}{c}{Storm}&\multicolumn{1}{c}{Heat}&\multicolumn{1}{c}{Flood}&\multicolumn{1}{c}{Drought}\\
\hline
Severity &       1.557\sym{***}&    1.333\sym{***}     &      0.0432\sym{***}               &   -0.0530   &   0.754\sym{***}&     0.00646   \\
                    &     (16.24)   &      (4.56) &   (13.34)  &   (-0.51)   &      (8.94)         &      (0.02)      \\
[1em]
Income per capita   &       0.240\sym{**} &      -0.401         &      -0.401\sym{***}&       0.184         &      -0.105\sym{**} &      -0.453\sym{***}\\
                    &      (3.18)         &     (-1.66)         &     (-6.68)         &      (0.91)         &     (-2.62)         &     (-3.87)         \\
[1em]
Tax revenue         &      -0.128\sym{***}&      -0.128         &     -0.0914\sym{***}&     -0.0563         &      -0.135\sym{***}&     -0.0683\sym{*}  \\
                    &     (-7.15)         &     (-1.55)         &     (-7.40)         &     (-1.54)         &    (-14.86)         &     (-2.52)         \\
[1em]
Income taxes        &     -0.0143\sym{**} &     -0.0150         &      0.0293\sym{***}&    -0.00115         &      0.0139\sym{***}&      0.0154         \\
                    &     (-2.65)         &     (-1.09)         &      (7.33)         &     (-0.08)         &      (4.49)         &      (1.43)         \\
[1em]
Democracy           &     -0.0576\sym{***}&       0.103         &     -0.0126         &      0.0619         &      0.0256\sym{**} &     -0.0490\sym{*}  \\
                    &     (-4.09)         &      (0.93)         &     (-1.02)         &      (1.08)         &      (2.67)         &     (-2.04)         \\
\hline
Observations        &        5882         &         316         &        7838         &        7835         &        8064         &        8064         \\
\hline\hline
\multicolumn{7}{l}{\footnotesize \textit{t} statistics in parentheses}\\
\multicolumn{7}{l}{\footnotesize \sym{*} \(p<0.05\), \sym{**} \(p<0.01\), \sym{***} \(p<0.001\)}\\
\end{tabular}
\end{table}

\begin{table}[htbp]\centering
\def\sym#1{\ifmmode^{#1}\else\(^{#1}\)\fi}
\caption{Regression results: ln(Fraction of people killed).\label{tab:regdis}}
\begin{tabular}{l*{6}{c}}
\hline\hline
                    &\multicolumn{1}{c}{Earthquake}&\multicolumn{1}{c}{Volcanoe}&\multicolumn{1}{c}{Storm}&\multicolumn{1}{c}{Heat}&\multicolumn{1}{c}{Flood}&\multicolumn{1}{c}{Drought}\\
\hline
Severity &       0.806\sym{**} &    1.274 &  0.00392   &    -0.101               &       0.303         &       0.111     \\
        &    (3.28)   & (.)  &   (1.00)    &   (-1.20)       &              (1.86)         &      (7.39)      \\
[1em]
Income per capita   &       0.133         &       0.913         &      -0.191\sym{*}  &       0.601\sym{***}&     -0.0537         &       0.492\sym{*}  \\
                    &      (0.58)         &         (.)         &     (-2.17)         &      (3.74)         &     (-0.82)         &     (46.05)         \\
[1em]
Tax revenue         &       0.101\sym{*}  &       2.231         &       0.134\sym{***}&      0.0889\sym{**} &      0.0749\sym{***}&       0.616\sym{**} \\
                    &      (2.09)         &         (.)         &      (7.49)         &      (2.77)         &      (4.88)         &    (285.50)         \\
[1em]
Income taxes        &     -0.0640\sym{***}&       0.102         &     -0.0207\sym{***}&     -0.0437\sym{***}&     -0.0341\sym{***}&      -0.167\sym{**} \\
                    &     (-3.70)         &         (.)         &     (-3.55)         &     (-3.86)         &     (-7.26)         &   (-116.57)         \\
[1em]
Democracy           &      0.0273         &       2.470         &      0.0725\sym{***}&      0.0461         &    -0.00271         &     -0.0274\sym{*}  \\
                    &      (0.65)         &         (.)         &      (4.25)         &      (0.81)         &     (-0.20)         &    (-30.93)         \\
\hline
Observations        &         171         &           6         &         487         &         111         &         547         &           7         \\
\hline\hline
\multicolumn{7}{l}{\footnotesize \textit{t} statistics in parentheses}\\
\multicolumn{7}{l}{\footnotesize \sym{*} \(p<0.05\), \sym{**} \(p<0.01\), \sym{***} \(p<0.001\)}\\
\end{tabular}
\end{table}

\begin{table}[htbp]\centering
\def\sym#1{\ifmmode^{#1}\else\(^{#1}\)\fi}
\caption{Regression results: ln(Fraction of people affected).\label{tab:regaff}}
\begin{tabular}{l*{6}{c}}
\hline\hline
                    &\multicolumn{1}{c}{Earthquake}&\multicolumn{1}{c}{Volcanoe}&\multicolumn{1}{c}{Storm}&\multicolumn{1}{c}{Heat}&\multicolumn{1}{c}{Flood}&\multicolumn{1}{c}{Drought}\\
\hline
Severity &       0.694\sym{**} &    0.232       &   0.0280\sym{***}    &   0.124      &    -0.195         &      -1.288   \\
        &      (3.33)         &   (0.51)          & (3.59) &  (0.06) &       (-0.64)         &     (-1.47)                  \\
[1em]
Income per capita   &       0.363         &      -1.374\sym{*}  &      -0.409\sym{*}  &      0.0641         &      -0.733\sym{***}&      -0.638\sym{*}  \\
                    &      (1.75)         &     (-2.38)         &     (-2.34)         &      (0.06)         &     (-6.68)         &     (-2.09)         \\
[1em]
Tax revenue         &       0.163\sym{***}&       0.240         &       0.131\sym{***}&       0.184         &       0.113\sym{***}&      0.0670         \\
                    &      (3.81)         &      (1.69)         &      (4.10)         &      (1.25)         &      (4.48)         &      (0.99)         \\
[1em]
Income taxes        &     -0.0692\sym{***}&      0.0382         &     -0.0523\sym{***}&      0.0145         &     -0.0172\sym{*}  &      0.0128         \\
                    &     (-5.02)         &      (1.00)         &     (-4.51)         &      (0.24)         &     (-2.27)         &      (0.48)         \\
[1em]
Democracy           &    -0.00371         &      0.0293         &     -0.0419         &       0.242         &     -0.0429         &      0.0189         \\
                    &     (-0.10)         &      (0.19)         &     (-1.26)         &      (0.96)         &     (-1.77)         &      (0.35)         \\
\hline
Observations        &         232         &          21         &         418         &          25         &         593         &          55         \\
\hline\hline
\multicolumn{7}{l}{\footnotesize \textit{t} statistics in parentheses}\\
\multicolumn{7}{l}{\footnotesize \sym{*} \(p<0.05\), \sym{**} \(p<0.01\), \sym{***} \(p<0.001\)}\\
\end{tabular}
\end{table}

\appendix \section{Additional results}

\begin{table}[h]\centering
\def\sym#1{\ifmmode^{#1}\else\(^{#1}\)\fi}
\caption{Regression results: Probability of people killed.\label{tab:logdis2}}
\begin{tabular}{l*{6}{c}}
\hline\hline
                    &\multicolumn{1}{c}{Earthquake}&\multicolumn{1}{c}{Volcanoe}&\multicolumn{1}{c}{Storm}&\multicolumn{1}{c}{Heat}&\multicolumn{1}{c}{Flood}&\multicolumn{1}{c}{Drought}\\
\hline
Severity &       2.077\sym{***}&  1.418\sym{*}  &  0.0476\sym{***}              &     -0.146\sym{*}      &   0.934\sym{***}&       0.342       \\
      &     (15.69)         &    (2.07)       &  (12.05)        &    (-2.00)      &     (9.16)         &      (0.37)            \\
[1em]
Income per capita   &      -0.431         &      -5.400         &       0.251         &       0.808         &       0.383         &      -2.654         \\
                    &     (-0.88)         &     (-1.02)         &      (0.79)         &      (1.31)         &      (1.29)         &     (-1.18)         \\
[1em]
Tax revenue         &     -0.0461         &       2.270         &     -0.0509         &    -0.00352         &     -0.0110         &       0.124         \\
                    &     (-0.62)         &      (1.92)         &     (-1.10)         &     (-0.06)         &     (-0.30)         &      (0.35)         \\
[1em]
Income taxes        &     -0.0194         &      -0.439         &    -0.00398         &    -0.00173         &      0.0238\sym{*}  &      0.0841         \\
                    &     (-1.19)         &     (-1.61)         &     (-0.38)         &     (-0.09)         &      (2.32)         &      (1.02)         \\
[1em]
Democracy           &      0.0898\sym{**} &       1.003         &      0.0145         &     0.00964         &      0.0441\sym{*}  &      -0.111         \\
                    &      (3.09)         &      (1.42)         &      (0.43)         &      (0.17)         &      (2.07)         &     (-0.38)         \\
\hline
Observations        &        3922         &         183         &        5961         &        4110         &        7284         &        1116         \\
\hline\hline
\multicolumn{7}{l}{\footnotesize \textit{t} statistics in parentheses}\\
\multicolumn{7}{l}{\footnotesize \sym{*} \(p<0.05\), \sym{**} \(p<0.01\), \sym{***} \(p<0.001\)}\\
\end{tabular}
\end{table}

\begin{table}[htbp]\centering
\def\sym#1{\ifmmode^{#1}\else\(^{#1}\)\fi}
\caption{Regression results: Probability of people affected.\label{tab:logdis12}}
\begin{tabular}{l*{6}{c}}
\hline\hline
                    &\multicolumn{1}{c}{Earthquake}&\multicolumn{1}{c}{Volcanoe}&\multicolumn{1}{c}{Storm}&\multicolumn{1}{c}{Heat}&\multicolumn{1}{c}{Flood}&\multicolumn{1}{c}{Drought}\\
\hline
Severity &       1.838\sym{***}&    2.102\sym{***}     &  0.0471\sym{***}  & -0.0315 &   0.934\sym{***}&      0.0605  \\
                    &     (16.07)        &    (4.53)     &    (11.86)  &   (-0.38)          &    (9.33)         &      (0.21)   \\
[1em]
Income per capita   &      -0.510         &      -1.753         &       0.957\sym{**} &       1.207         &       0.994\sym{***}&      -1.767         \\
                    &     (-1.17)         &     (-0.84)         &      (2.83)         &      (1.04)         &      (3.39)         &     (-1.77)         \\
[1em]
Tax revenue         &     -0.0433         &      0.0702         &      0.0136         &      -0.143         &    -0.00935         &      -0.205         \\
                    &     (-0.67)         &      (0.32)         &      (0.33)         &     (-1.49)         &     (-0.28)         &     (-1.79)         \\
[1em]
Income taxes        &     -0.0173         &     -0.0359         &    -0.00704         &      0.0281         &      0.0172         &      0.0246         \\
                    &     (-1.20)         &     (-0.72)         &     (-0.65)         &      (0.73)         &      (1.71)         &      (0.70)         \\
[1em]
Democracy           &      0.0861\sym{**} &       0.172         &     -0.0450         &       0.109         &      0.0535\sym{*}  &       0.167         \\
                    &      (3.24)         &      (0.96)         &     (-1.32)         &      (0.77)         &      (2.50)         &      (1.90)         \\
\hline
Observations        &        4147         &         192         &        6556         &        3096         &        7296         &        3108         \\
\hline\hline
\multicolumn{7}{l}{\footnotesize \textit{t} statistics in parentheses}\\
\multicolumn{7}{l}{\footnotesize \sym{*} \(p<0.05\), \sym{**} \(p<0.01\), \sym{***} \(p<0.001\)}\\
\end{tabular}
\end{table}

\begin{table}[h]\centering
\def\sym#1{\ifmmode^{#1}\else\(^{#1}\)\fi}
\caption{Regression results: ln(Fraction of people killed).\label{tab:regdis2}}
\begin{tabular}{l*{6}{c}}
\hline\hline
                    &\multicolumn{1}{c}{Earthquake}&\multicolumn{1}{c}{Volcanoe}&\multicolumn{1}{c}{Storm}&\multicolumn{1}{c}{Heat}&\multicolumn{1}{c}{Flood}&\multicolumn{1}{c}{Drought}\\
\hline
Severity &       0.948\sym{***}&   1.274  &    0.0139\sym{***}   &   -0.0611  &           -0.0252         &   -3.06e-13        \\
                    &      (3.53)         & (.)    &    (4.13)  &     (-0.62)     &       (-0.15)         &         (.)          \\
[1em]
Income per capita   &       1.477         &       0.913         &      -2.033\sym{***}&       0.486         &      -1.575\sym{***}&       0.530         \\
                    &      (1.23)         &         (.)         &     (-5.55)         &      (0.45)         &     (-4.35)         &         (.)         \\
[1em]
Tax revenue         &       0.338         &       2.231         &      0.0199         &     -0.0996         &     -0.0590         &       0.608         \\
                    &      (1.84)         &         (.)         &      (0.37)         &     (-0.98)         &     (-1.25)         &         (.)         \\
[1em]
Income taxes        &     -0.0209         &       0.102         &      0.0281\sym{*}  &     -0.0438         &      0.0251\sym{*}  &      -0.168         \\
                    &     (-0.50)         &         (.)         &      (2.32)         &     (-1.16)         &      (2.08)         &         (.)         \\
[1em]
Democracy           &     -0.0178         &       2.470         &      0.0300         &       0.113         &   -0.000558         &     -0.0346         \\
                    &     (-0.26)         &         (.)         &      (0.69)         &      (0.96)         &     (-0.02)         &         (.)         \\
[1em]
\hline
Observations        &         171         &           6         &         487         &         111         &         547         &           7         \\
\hline\hline
\multicolumn{7}{l}{\footnotesize \textit{t} statistics in parentheses}\\
\multicolumn{7}{l}{\footnotesize \sym{*} \(p<0.05\), \sym{**} \(p<0.01\), \sym{***} \(p<0.001\)}\\
\end{tabular}
\end{table}

\begin{table}[h]\centering
\def\sym#1{\ifmmode^{#1}\else\(^{#1}\)\fi}
\caption{Regression results: ln(Fraction of people affected).\label{tab:regaff2}}
\begin{tabular}{l*{6}{c}}
\hline\hline
                    &\multicolumn{1}{c}{Earthquake}&\multicolumn{1}{c}{Volcanoe}&\multicolumn{1}{c}{Storm}&\multicolumn{1}{c}{Heat}&\multicolumn{1}{c}{Flood}&\multicolumn{1}{c}{Drought}\\
\hline
Severity &       0.794\sym{***}&  0.685    &   0.0383\sym{***}   &  6.085    &     -0.321         &      -1.168     \\
            &      (3.56)         &      (0.94) & (5.08)  &   (0.32)  &         (-0.95)         &     (-1.12)     \\
[1em]
Income per capita   &       3.334\sym{**} &     -0.0251         &      -2.273\sym{**} &      -17.87\sym{*}  &      -1.410         &       2.788         \\
                    &      (3.19)         &     (-0.01)         &     (-2.60)         &     (-2.70)         &     (-1.85)         &      (1.17)         \\
[1em]
Tax revenue         &       0.178         &       0.302         &       0.147         &       0.321         &     -0.0424         &     -0.0525         \\
                    &      (1.14)         &      (1.28)         &      (1.24)         &      (0.46)         &     (-0.50)         &     (-0.12)         \\
[1em]
Income taxes        &     -0.0621         &     0.00302         &     -0.0408         &       0.291         &     -0.0161         &   -0.000184         \\
                    &     (-1.62)         &      (0.03)         &     (-1.46)         &      (0.91)         &     (-0.62)         &     (-0.00)         \\
[1em]
Democracy           &      0.0557         &      -0.128         &      0.0469         &       2.461         &     0.00757         &     0.00298         \\
                    &      (0.93)         &     (-0.35)         &      (0.46)         &      (0.27)         &      (0.15)         &      (0.01)         \\
[1em]
\hline
Observations        &         232         &          21         &         418         &          25         &         593         &          55         \\
\hline\hline
\multicolumn{7}{l}{\footnotesize \textit{t} statistics in parentheses}\\
\multicolumn{7}{l}{\footnotesize \sym{*} \(p<0.05\), \sym{**} \(p<0.01\), \sym{***} \(p<0.001\)}\\
\end{tabular}
\end{table}

\begin{table}[htbp]\centering \footnotesize
\def\sym#1{\ifmmode^{#1}\else\(^{#1}\)\fi}
\caption{Tobit regression results: Fraction of people killed.\label{tab:tobkill}}
\begin{tabular}{l*{6}{c}}
\hline\hline
                    &\multicolumn{1}{c}{Earthquake}&\multicolumn{1}{c}{Volcano}&\multicolumn{1}{c}{Storm}&\multicolumn{1}{c}{Heat}&\multicolumn{1}{c}{Flood}&\multicolumn{1}{c}{Drought}\\
\hline
Severity &   0.0000790\sym{***}& 0.000000401\sym{*}   & 0.000000613\sym{***} &  3.70e-08  &   0.00000235\sym{***}&    7.72e-08   \\
                    &     (11.81)    &  (1.98)   &   (10.64) & (1.11)  & (7.80)         &      (0.03)     \\
[1em]
Income per capita   &  0.00000928\sym{*}  &-0.000000170         & -0.00000717\sym{***}&-0.000000129         &-0.000000468\sym{***}& -0.00000274\sym{*}  \\
                    &      (2.24)         &     (-0.95)         &     (-7.09)         &     (-1.75)         &     (-3.42)         &     (-2.10)         \\
[1em]
Tax revenue         & -0.00000523\sym{***}&    5.34e-08         & -0.00000163\sym{***}&   -3.11e-08         &-0.000000418\sym{***}&    1.79e-08         \\
                    &     (-5.32)         &      (1.25)         &     (-7.97)         &     (-1.83)         &    (-12.85)         &      (0.09)         \\
[1em]
Income taxes        & -0.00000101\sym{**} &   -1.70e-09         & 0.000000480\sym{***}&    1.31e-08\sym{*}  &    3.82e-08\sym{***}& 0.000000217\sym{*}  \\
                    &     (-3.22)         &     (-0.18)         &      (7.02)         &      (2.46)         &      (3.64)         &      (2.02)         \\
[1em]
Democracy           & -0.00000216\sym{**} & 0.000000100         & 0.000000280         &    6.60e-08         &    5.44e-08         &-0.000000378         \\
                    &     (-2.74)         &      (0.99)         &      (1.38)         &      (1.34)         &      (1.80)         &     (-1.62)         \\
[1em]
\hline
Observations        &        5882         &         316         &        7838         &        7741         &        8064         &        8064         \\
\hline\hline
\multicolumn{7}{l}{\footnotesize \textit{t} statistics in parentheses}\\
\multicolumn{7}{l}{\footnotesize \sym{*} \(p<0.05\), \sym{**} \(p<0.01\), \sym{***} \(p<0.001\)}\\
\end{tabular}
\end{table}

\begin{table}[htbp]\centering
\def\sym#1{\ifmmode^{#1}\else\(^{#1}\)\fi}
\caption{Tobit regression results: Fraction of people affected.\label{tab:tobaff}}
\begin{tabular}{l*{6}{c}}
\hline\hline
                    &\multicolumn{1}{c}{Earthquake}&\multicolumn{1}{c}{Volcano}&\multicolumn{1}{c}{Storm}&\multicolumn{1}{c}{Heat}&\multicolumn{1}{c}{Flood}&\multicolumn{1}{c}{Drought}\\
\hline
Severity &      0.0104\sym{***}&  0.0119\sym{***}    &  0.000576\sym{***}  &  -0.00131   &  0.00790\sym{***}&     -0.0143    \\
                    &     (12.57)         &   (3.77)      &  (12.32)   &  (-0.47)  &        (7.98)         &     (-0.37)     \\
[1em]
Income per capita   &     0.00201\sym{***}&    -0.00498\sym{*}  &    -0.00575\sym{***}&     0.00333         &    -0.00169\sym{***}&     -0.0559\sym{***}\\
                    &      (3.53)         &     (-1.98)         &     (-7.03)         &      (0.73)         &     (-3.80)         &     (-3.37)         \\
[1em]
Tax revenue         &   -0.000600\sym{***}&   -0.000561         &   -0.000717\sym{***}&    -0.00110         &    -0.00119\sym{***}&    -0.00750\sym{*}  \\
                    &     (-4.80)         &     (-0.75)         &     (-4.66)         &     (-1.35)         &    (-11.84)         &     (-2.28)         \\
[1em]
Income taxes        &   -0.000125\sym{**} &  -0.0000639         &    0.000331\sym{***}&   0.0000682         &    0.000149\sym{***}&     0.00210         \\
                    &     (-3.15)         &     (-0.46)         &      (6.05)         &      (0.22)         &      (4.42)         &      (1.59)         \\
[1em]
Democracy           &   -0.000374\sym{***}&    0.000720         &  -0.0000608         &     0.00136         &    0.000187         &    -0.00524         \\
                    &     (-3.49)         &      (0.80)         &     (-0.38)         &      (1.09)         &      (1.85)         &     (-1.69)         \\
\hline
Observations        &        5882         &         316         &        7838         &        7835         &        8064         &        8064         \\
\hline\hline
\multicolumn{7}{l}{\footnotesize \textit{t} statistics in parentheses}\\
\multicolumn{7}{l}{\footnotesize \sym{*} \(p<0.05\), \sym{**} \(p<0.01\), \sym{***} \(p<0.001\)}\\
\end{tabular}
\end{table}

\end{document}